\documentclass[a4paper,11pt]{article}
\pdfoutput=1 

\usepackage[usenames,dvipsnames,svgnames,table]{xcolor} 
\usepackage{jcapmod}
\usepackage{verbatim}
\usepackage{tikz}

\usepackage[T1]{fontenc} 
\definecolor{lightgray}{gray}{0.9}
\definecolor{Amber}{rgb}{1.0, 0.75, 0.0}
\definecolor{blizzardblue}{rgb}{0.67, 0.9, 0.93}

\usepackage{float}
\usepackage{graphicx}
\usepackage{hyperref}
\usepackage{amsmath}
\usepackage{amssymb}
\usepackage{microtype}
\usepackage{amsbsy}
\usepackage[usenames,dvipsnames,table]{xcolor}
\usepackage[capitalise]{cleveref}
\usepackage{breakurl}
\usepackage{csquotes}
\usepackage{lmodern}
\usepackage{bm}
\usepackage{dsfont}
\usepackage{soul}

\pdfpageheight\paperheight
\pdfpagewidth\paperwidth

\renewcommand*{\vec}[1]{\bm{#1}}

\newcommand*{\mat}[1]{\bm{\mathsf{#1}}}

\newcommand{\dderiv}{\mathrm{d}}
\newcommand{\identity}{\ensuremath{\mathds{1}}}

\newcommand\integers{\mathbb{Z}}
\newcommand\reals{\mathbb{R}}

\newcommand*{\Mod}[1]{\,(\mathrm{mod}\ #1)}
\newcommand{\dLSS}{\ensuremath{d_{\mathrm{LSS}}}}

\allowdisplaybreaks

\makeatletter
\DeclareRobustCommand{\rcite}[1]{%
  \rcite@aux#1,\@nil{#1}%
}
\def\rcite@aux#1,#2\@nil#3{%
  \if\relax#2\relax
    Ref.~\cite{#3}%
  \else
    Refs.~\cite{#3}%
  \fi
}
\makeatother

\subheader{IFT-UAM/CSIC-24-118}

\title{Cosmic topology. Part Ic. Limits on lens spaces from circle searches}

\author[a]{Samanta Saha,}
\author[a]{Craig J. Copi,}
\author[a]{Glenn D. Starkman,}
\author[b,c,d]{Stefano Anselmi,}
\author[e]{Javier Carr\'on Duque,}
\author[e]{Mikel Martin Barandiaran,}
\author[e,a,f]{Yashar Akrami,}
\author[a]{Fernando Cornet-Gomez,}
\author[f]{Andrew H. Jaffe,}
\author[g]{Arthur Kosowsky,}
\author[a]{Deyan P. Mihaylov,}
\author[h]{Thiago S. Pereira,}
\author[a]{Amirhossein Samandar,}
\author[a]{and Andrius Tamosiunas}

\collaboration{(COMPACT Collaboration)}
\collaborationImg{\includegraphics[width = 0.1\textwidth]{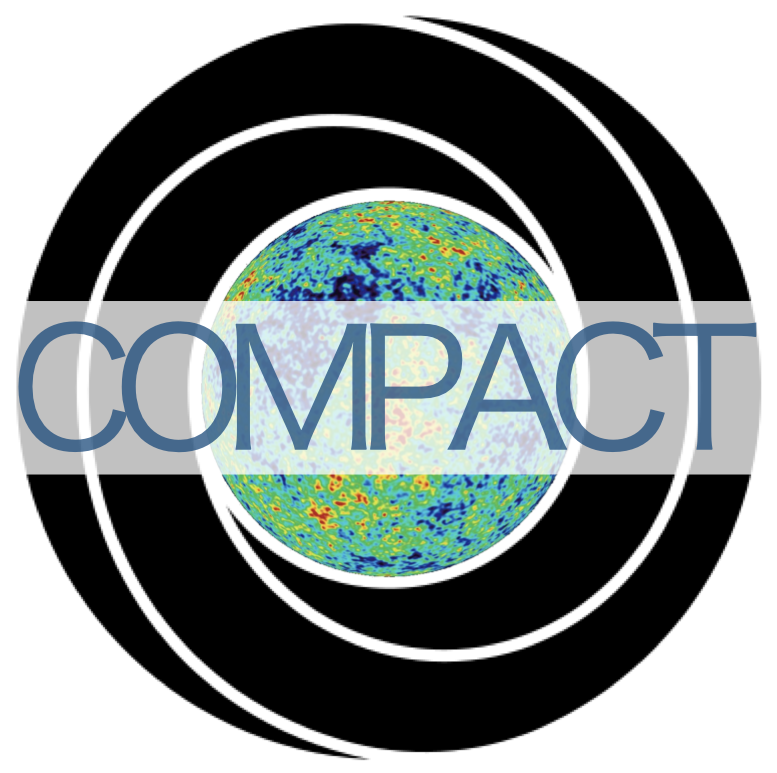}}

\affiliation[a]{CERCA/ISO, Department of Physics, Case Western Reserve University, 10900 Euclid Avenue, Cleveland, Ohio 44106, USA}
\affiliation[b]{INFN, Sezione di Padova, via Marzolo 8, I-35131 Padova, Italy}
\affiliation[c]{Dipartimento di Fisica e Astronomia ``G. Galilei'', Universit\`a degli Studi di Padova, via Marzolo 8, I-35131 Padova, Italy}
\affiliation[d]{Laboratoire Univers et Th\'eories, Observatoire de Paris, Universit\'e PSL, Universit\'e Paris Cit\'e, CNRS, F-92190 Meudon, France}
\affiliation[e]{Instituto de F\'isica Te\'orica (IFT) UA
M-CSIC, C/ Nicol\'as Cabrera 13-15, Campus de Cantoblanco UAM, 28049 Madrid, Spain}
\affiliation[f]{Astrophysics Group \& Imperial Centre for Inference and Cosmology, Department of Physics, Imperial College London, Blackett Laboratory, Prince Consort Road, London SW7 2AZ, United Kingdom}
\affiliation[g]{Department of Physics and Astronomy, University of Pittsburgh, Pittsburgh, Pennsylvania 15260, USA}
\affiliation[h]{Departamento de F\'{i}sica, Universidade Estadual de Londrina,
Rod. Celso Garcia Cid, Km 380, 86057-970, Londrina, Paran\'a, Brazil}

\emailAdd{samanta.saha@case.edu}
\emailAdd{craig.copi@case.edu}
\emailAdd{glenn.starkman@case.edu}
\emailAdd{stefano.anselmi@pd.infn.it}
\emailAdd{javier.carron@csic.es}
\emailAdd{mikel.martin@uam.es}
\emailAdd{yashar.akrami@csic.es}
\emailAdd{fernando.cornetgomez@case.edu}
\emailAdd{a.jaffe@imperial.ac.uk}
\emailAdd{kosowsky@pitt.edu}
\emailAdd{deyan.mihaylov@case.edu}
\emailAdd{tspereira@uel.br}
\emailAdd{amirhossein.samandar@case.edu}
\emailAdd{andrius.tamosiunas@case.edu}

\date{\today}

\abstract{
    Cosmic microwave background (CMB) temperature and polarization observations indicate that in the best-fit $\Lambda$ Cold Dark Matter  model of the Universe, the local geometry is consistent with at most a small amount of positive or negative curvature, i.e., $\vert\Omega_K\vert\ll1$.
    However, whether the geometry is flat ($E^3$), positively curved ($S^3$) or negatively curved ($H^3$), there are many possible topologies.
    Among the topologies of $S^3$ geometry, the lens spaces $L(p,q)$, where $p$ and $q$ ($p>1$ and $0<q<p$) are positive integers, are quotients of the covering space of $S^3$ (the three-sphere) by $\integers_p$, the cyclic group of order $p$.
    We use the absence of any pair of circles on the CMB sky with matching patterns of temperature fluctuations to establish constraints on  $p$ and $q$  as a function of the curvature scale that are considerably stronger than those previously asserted for most values of $p$ and $q$.
    The smaller the value of $\vert\Omega_K\vert$, i.e., the larger the curvature radius, the larger the maximum allowed value of $p$.
    For example, if $\vert\Omega_K\vert\simeq 0.05$ then $p\leq 9 $, while if $\vert\Omega_K\vert\simeq 0.02$, $p$ can be as high as 24. 
    Future work will extend these constraints to a wider set of $S^{3}$ topologies.
}


\begin{document}
\maketitle
\flushbottom

\section{Introduction}
\label{sec:introduction}
According to the general theory of relativity, the local geometry of spacetime is a solution of the Einstein field equations, a set of coupled, non-linear, local partial differential equations \cite{Einstein:1916vd}. 
Assuming that appropriate spatial slices of that spacetime and its contents are homogeneous and isotropic, the local four-geometry will be described by one of the three Friedmann-Lemaître-Robertson-Walker (FLRW) metrics, i.e., one of the three isotropic, constant-curvature geometries---flat (zero spatial curvature) Euclidean geometry $E^3$, positive spatial curvature $S^3$, and negative spatial curvature $H^3$---with a time-dependent scale factor. 

The metric, however, characterizes only the local geometry---one must independently specify the global structure of the spacetime. 
It is conventionally assumed  that the spacetime manifold is time cross the covering space\footnote{
    Though it can be confusing, we will adopt the usual custom of referring both to the local flat, positively curved, and negatively curved geometries and their covering spaces as, respectively, $E^3$, $S^3$, and $H^3$.
}
of one of those three homogeneous, isotropic geometries---i.e., infinite flat 3-space (the covering space of $E^3$), the 3-sphere $S^3$, or the 3-dimensional pseudosphere $H^3$.
However, while these covering spaces are the manifolds that globally preserve the isometry group of each local geometry, each homogeneous and isotropic 3-geometry admits many possible topologies, with at least one real parameter to specify the manifold with that topology.

There are thus many possible 3-manifolds other than the three covering spaces that can accommodate FLRW cosmology.
They are distinguished from the covering spaces in lacking globally the full isotropy\footnote{
    Except for the projective space $S^3/\integers_2\cong L(2,1)$, which is the only multiply-connected FLRW manifold that is both globally homogeneous and isotropic.
}
(and usually parity and homogeneity) of their local geometries.
This global isotropy breaking, parity breaking, or homogeneity breaking  has the potential to be reflected in the properties of fluctuations around the FLRW background, and thus in the statistical properties of cosmological observables. In recent work \cite{COMPACT:2024cud}, we explored how symmetry breaking in non-trivial topologies affects cosmic microwave background (CMB) polarization correlation matrices.

In this paper, we consider the possible spacetimes with $S^3$ spatial geometry, and place new limits on one class of allowed $S^3$ topologies---\emph{lens spaces}---as a function of the curvature scale.

While 3-space on large scales is reasonably homogeneous and isotropic, there is evidence from CMB temperature fluctuations of multiple ``large-angle'' anomalies.
Together, these amount to evidence in excess of $5\sigma$ equivalent significance against statistical isotropy \cite{Jones:2023ncn}, mostly on scales larger than the horizon size at the time of the last scattering of the CMB photons (see, e.g., \rcite{Planck:2013lks,Schwarz:2015cma,Planck:2015igc,Planck:2019evm,Abdalla:2022yfr} for reviews.)
One of the very few potential physical explanations for this large-scale anisotropy is non-trivial spatial topology, though no specific manifold has yet been identified that explains the observed isotropy violation.

The possibility of non-trivial spatial topology has been considered since at least as far back as Einstein's initial $S^3$
cosmological model \cite{Einstein:1917ce} at which time de Sitter remarked \cite{deSitter:1917zz} that the projective 3-sphere, which has the same local geometry but half the 3-volume, was to be preferred.
Ever since, there have been cosmologists working to develop observational tests of cosmic topology.
Several approaches have been considered, including cosmic crystallography (the search for topological ``clones'') \cite{Lehoucq:1996qe}, CMB 
matched circle pairs (a.k.a.\ ``circles in the sky'') \cite{Cornish:1997ab,Cornish:1997rp,Cornish:2003db} which is essentially the search for topological clones in CMB maps, and CMB Bayesian likelihood comparison \cite{Kunz:2005wh,Planck:2013okc,Planck:2015gmu}. 
More recently, the general set of eigenmodes, the correlation matrices, and the detectability of the orientable Euclidean manifolds have been studied in \rcite{COMPACT:2022gbl, COMPACT:2022nsu, COMPACT:2023rkp, COMPACT:2024cud}. Similarly, various machine learning techniques have been explored as tools for detecting signatures of non-trivial topology in harmonic space \cite{COMPACT:2024dqe}.

While, in principle, Bayesian likelihood comparison is the most powerful technique because it uses all the available data,  it suffers the disadvantage that the likelihood must be computed for each allowed topology of each allowed 3-geometry, for every value of the  real parameters that specify the 3-manifolds of a given topology, and for every distinguishable position and orientation of the observer within the manifold \cite{Kunz:2005wh}. Moreover, that scan over the space of possible 3-manifolds should, at least in principle, be done at the same time as the scan over the space of parameters that specify the background cosmology.
With a countable infinity of possible topologies, up to six additional parameters that specify the manifold (e.g., the lengths of the sides of a torus in $E^3$) and up to six parameters specifying the position and orientation of the observer within a manifold that breaks isotropy and homogeneity, in addition to the seven cosmological parameters, it is no surprise that the full Bayesian likelihood search has yet to be attempted. 
On the other hand, a search using the circles-in-the-sky method is agnostic to the cosmological parameters, agnostic to the local 3-geometry---so long as anisotropies in the 3-geometry are sufficiently small---and agnostic to the values of the topological parameters within the domain of validity of the search.   

In a topologically non-trivial spatial 3-manifold, points in the covering space of the local geometry are identified if they are related by any element of some discrete group of spatial transformations (a discrete, freely acting, subgroup of the isometry group of the local geometry).
We call any such pair of identified points in the covering space  ``clones'' (or sometimes ``topological clones''). It is important to realize that these clones have no independent existence, but they are often a convenient way to visualize or calculate the consequences of non-trivial topology.

The circles signature rests on two fundamental observations.
First, the CMB photons have all been traveling through the Universe since very nearly the same time, i.e., since recombination of the primordial plasma, and therefore the CMB that any observer detects comes from a sphere centered on them---their last-scattering surface (LSS)\@.
Second, for every one of our clones that is closer to us than the diameter $\dLSS$ of the LSS, our LSS intersects with ``their'' LSS and the intersection of those two spheres is a circle. 
This circle is a locus of points visible to us (i.e., to ourselves and our clone) in two distinct directions on the sky.\footnote{
    This description pretends that the LSS has zero thickness. The actual LSS has a finite thickness, and so the self-intersection of the LSS is a finite-volume circular tube with a complicated cross-sectional profile.}
So long as that circle is large enough, we would be able to identify the tight correlation of CMB temperature fluctuations around the two matched circles as statistically anomalous, and so detect non-trivial topology.

This search for matched-circle pairs was performed in full generality on the Wilkinson Microwave Anisotropy Probe (WMAP) full-sky temperature map and no statistically significant matches were found \cite{Cornish:2003db,Vaudrevange:2012da,ShapiroKey:2006hm}. 
The search was repeated on {\it Planck} 2013 maps with identical results \cite{Starkman_Priv_Comm}, but not reported.
A limited search was conducted by the {\it Planck} team, again with negative outcome \cite{Planck:2013okc,Planck:2015gmu}.
In all cases, the smallest circle that could be ruled out depended on the required false negative and positive rates. 
However, both are steep functions of the radius of the circle for small circle size.\footnote{
    A more subtle question is how well to trust limits on those circles that are small enough to lie entirely or mostly within the usual foreground masks that are applied to full-sky CMB temperature maps.  
    We reserve such questions for a future paper revisiting the statistical details of the circle searches.} 
If the topology scale of the Universe were sufficiently small to produce matched circles on the CMB sky, they would appear in both temperature and polarization. Therefore, future full-sky observations of the CMB polarization are expected to provide a complementary verification of the lack of circles; this probe is valuable as it is sensitive to different systematics and foregrounds, but it has limited discovery power as CMB temperature already constrains most of the parameter space accessible by circle searches.

The negative conclusion of these searches is that the length of the shortest path from us to our nearest clone, i.e., the shortest path around the Universe \emph{through us}, must be greater than $f_\mathrm{O} \dLSS$.
The reported value of $f_\mathrm{O}$ is $0.985$ at 95\% confidence level. 
The task remains to translate this generic limit on the distance to our nearest clone into a constraint on model parameters.   
In \rcite{COMPACT:2022nsu}, this was done for orientable manifolds admitting homogeneous Euclidean geometry $E^3$; this will be complemented in a future paper on non-orientable $E^3$ manifolds.
Here, we begin the same task for manifolds admitting a homogeneous $S^3$ local geometry by considering the lens spaces $L(p,q)$: the quotient spaces of the spherical manifolds by the cyclic group $\integers_p$.
A future paper will consider the other $S^3$ manifolds, 
and still other papers will in turn consider the six other Thurston geometries \cite{Thurston1982ThreeDM} as each presents its own particular challenges.

This work is by no means the first attempt to constrain topologies of $S^3$.
A detailed construction and complete classification of all 3-dimensional spherical manifolds was given by Gausmann et al.~\cite{Gausmann:2001aa}.
They also discussed the likelihood of detectability of spherical topologies by crystallographic methods, as a function of cosmological parameters. 
Gomero et al.~\cite{Gomero:2001gq} considered which hyperbolic and spherical manifolds were excluded by observations, however, this was before the considerable progress made using WMAP data and then {\it Planck} data, and in particular they could not include constraints from matched-circle searches 
\cite{Cornish:2003db,ShapiroKey:2006hm,Vaudrevange:2012da,Planck:2013okc,Planck:2015gmu}.
In the same period, Uzan et al.~\cite{Uzan:2003ea} studied CMB anisotropies in $S^3$ manifolds, focusing on the lens spaces $L(p,1)$.
Suppression of low-$\ell$ anisotropies in inhomogeneous lens spaces $L(p,q)$, especially with $p=8$, was studied by Aurich et al.~\cite{Aurich:2011wb}, followed by exploration of the specific $L(p, q=p/2 - 1)$ with $p\Mod{4}=0$ and prism spaces \cite{Aurich:2012aa}.
In \rcite{Aurich:2012sp}, the authors surveyed lens spaces with $p \leq 72$ and concluded that $L(p,q)$ with $q\approx 0.28p$ and $q\approx 0.38p$ display strong suppression of CMB fluctuations on angular scales $\theta \geq 60^\circ$ compared with  the covering  space.

It is important to note that, for all $p$ and all allowed values of $q\Mod{p}>1$, $L(p,q)$ is statistically inhomogeneous, i.e., the statistical properties of CMB anisotropies (and other observables) depend on the CMB observer's position. 
This is because translation invariance is broken by the requisite boundary conditions---for example, the length of the shortest closed geodesic varies with location.
The $q=1$ lens spaces are the rare exception.
This means that any exploration of constraints on lens spaces must vary not only $p$, $q$, and the Ricci scalar, but also the location  of the observer. Similarly both isotropy and parity are violated statistically; and so the orientation of the observer and the handedness of their coordinate system matter.

In this paper, we systematically explore constraints on $L(p,q)$, based on the fact that matched-circle pairs in the CMB temperature sky have not been detected.
This, and a new analysis of clone separations in $L(p,q)$,  will allow us to considerably strengthen the previous limits on these spaces obtained in \rcite{Gomero:2001gq}.

This paper is organized as follows. We provide a brief review of the $S^3$ geometry and lens spaces $L(p,q)$ in \cref{sec:3-sphere}.
The background for the application of circle searches to lens spaces is given in \cref{sec:circle_searches}.
To connect the circle searches to observational constraints requires understanding of $S^3$ cosmological models, which is presented in \cref{sec:cosmology}.
In \cref{sec:constraints}, we relate the non-detection of CMB matched-circle pairs to constraints on the parameters of the lens spaces and gives a strong condition on the detectability of the lens spaces as possible topologies of the Universe.
We summarize the paper and conclude in \cref{sec:conclusions}.

The GitHub repository associated with this study is publicly available at \url{https://github.com/CompactCollaboration}. Codes will be deposited there as publicly usable versions become available.

\section{3-sphere and lens spaces }
\label{sec:3-sphere}
We begin with a short introduction to $S^3$ and, in particular, the lens spaces.
Some details not pertinent to limits from circle searches are included to provide a more complete introduction and a foundation for future studies.

There are various ways to discuss the 3-sphere, $S^3$.
We will describe it in terms of its natural embedding in 4-dimensional Euclidean space $E^4$ as the set of points $\{(x_0, x_1, x_2, x_3) \mid x_0^2+x_1^2+x_2^2+x_3^2=R_c^2\}$.
For simplicity we will typically work in units of the curvature scale, i.e., we set $R_c=1$.

The 3-sphere $S^3$ is both this simply-connected space (i.e., any closed loop on this 3-sphere can be smoothly contracted to a point) and the geometry induced on this 3-sphere.
In this representation, it is manifest that the isometry group of $S^3$ is $O(4)$---the rotations and reflections in four dimensions.  
The topologically non-trivial manifolds with  $S^3$ geometry are quotients of the 3-sphere by any freely acting discrete subgroup of the full isometry group.\footnote{
    A group is freely acting if the only group element that takes any point (on the 3-sphere) to itself is the identity.
} 
For $S^3$ the freely acting discrete subgroups consist of only rotations, i.e., they are all discrete subgroups of $SO(4)$, none of the freely acting isometries are parity-reversing.
More simply put, one covers the 3-sphere with a finite number of identical tiles that are related to one another by (a finite set of) $SO(4)$ elements, such that no point on the edge of one tile touches the identical point on a neighboring tile, and the tiles share only edges.
There are a countably infinite number of such discrete subgroups of $SO(4)$.
Threlfall and Seifert \cite{Threlfall1931} gave the first complete classification of these spherical 3-manifolds. 
Another classification method, using quaternions, can be found in \rcite{Thurston1982ThreeDM}. 
In the cosmological context, a detailed and complete classification of all the spherical manifolds can be found in  \rcite{Gausmann:2001aa}.

In this paper, we focus on lens spaces, quotient spaces of $S^3$ of the form $S^3/\integers_p$, where $\integers_p$ is the cyclic group of order $p$ (considered as a subgroup of $SO(4)$).
There are multiple distinct actions of $\integers_p$ on $S^3$ that give distinct spaces labeled by a second integer parameter $q$, where $p$ and $q$ are relatively prime and  $0<q<p$ \cite{Thurston1982ThreeDM}. The group of each of these actions has $p$ elements, $\mat{R}^j_{pq}$ (with $j \in \{0,\ldots,p-1 \}$), acting on a point $\vec{x}\in S^3$  by 
\begin{equation}
    \vec{x}' = \left(\mat{R}_{pq}\right)^j \vec{x} \equiv  \mat{R}_{pq}^j \vec{x}\,.
\end{equation}
  
One particularly useful representation of $\mat{R}^j_{pq}$ is as a rotation in $E^4$ separated into rotations in two orthogonal planes,
\begin{equation}
    \label{eqn:Rjpq}
    \mat{R}^j_{pq}=
    \begin{pmatrix}
        \cos(2\pi j/p)  & -\sin(2\pi j/p) & 0 & 0 \\ 
        \sin(2\pi j/p)  & \hphantom{-}\cos(2\pi j/p) & 0 & 0 \\
        0 & 0 &\cos(2\pi jq/p)  & -\sin(2\pi jq/p) \\
        0 & 0 &\sin(2\pi jq/p)  & \hphantom{-}\cos(2\pi jq/p) \\
    \end{pmatrix} \,.
\end{equation}
This contains a rotation by $(2\pi/p)j$ in the $x_0$-$x_1$ plane simultaneously with a rotation in the $x_2$-$x_3$ plane by $(2\pi/p) (jq\Mod{p})$.
Note that $\mat{R}^j_{pq}$ keeps both $x_0^2+x_1^2$ and $x_2^2+x_3^2$ unchanged---this will prove crucial for understanding the limit we will place on $L(p,q)$.

Clearly $\mat{R}_{pq}^p=\mat{R}_{pq}^0\equiv\identity$,
while $\mat{R}_{pq}\equiv \mat{R}^1_{pq}$ is a generator of the group.\footnote{
    It would seem that one could generalize \eqref{eqn:Rjpq} by replacing $j/p$ by $q/p$ in the $x_0$-$x_1$ block and $jq/p$ by $jq'/p'$ in the $x_2$-$x_3$ block for a wider variety of integers $p,p',q$ and $q'$.
    However, one can easily show that all combinations other than those given by \eqref{eqn:Rjpq} are either not freely acting or equivalent to one of those already considered.
}
An object at a location $\vec{x}^{(0)}$ would thus have $p-1$ distinct clones,
\begin{equation}
    \label{eqn:cloneposn}
    \vec{x}^{(j)} = \mat{R}^j_{pq} \vec{x}^{(0)}\,, \qquad j\in\{1,\ldots,p-1\}\,.
\end{equation}

For each choice of the value of $p$, it appears at first sight that there are up to $p-1$ distinct $q$ values defining different lens spaces $L(p,q)$ for $q\in\{1, \ldots, p-1\}$.
Note that $\mat{R}_{p0}$ is not freely acting, and that, if $q>p$, $\mat{R}_{pq} = \mat{R}_{p(q\Mod{p})}$, so we can indeed limit the analysis to $0<q<p$.
In truth, not all of these values are allowed---some are not freely acting---and not all of them are distinct.  
However, let us first understand the role of $q$.
To do so, consider  the pattern of clones yielded by \eqref{eqn:cloneposn} given the representation of $\mat{R}_{pq}$ in \eqref{eqn:Rjpq}. 

We can label the clones of any point by the two integers $j$ and $j'\equiv jq\Mod{p}$---for a fixed $p$ and $q$ there will be $p-1$ clones of the initial point.
Taking $p=7$ as an example, for $q=1$, the rotation $\mat{R}_{71}$ takes $(j,j')$ to $((j+1)\Mod{7},(j'+1)\Mod{7})$. Starting at $(j,j')=(0,0)$, repeated applications of $\mat{R}_{71}$ gives the $(j,j')$ sequence
\begin{equation}
(0,0)\to(1,1)\to(2,2)\to(3,3)\to(4,4)\to(5,5)\to(6,6)\to(0,0).\label{eqn:R71-pattern}
\end{equation} 
This represents the 6 clones of a point labeled by $(0,0)$ in $L(7,1)$.

Similarly, for $q=2$, the rotation $\mat{R}_{72}$ takes $(j,j')\to((j+1)\Mod{7} , (j'+2)\Mod{7})$,
so starting at $(0,0)$ repeated applications of $\mat{R}_{72}$ takes
\begin{equation}
(0,0)\to(1,2)\to(2,4)\to(3,6)\to(4,1)\to(5,3)\to(6,5)\to(0,0).
\label{eqn:R72-pattern}
\end{equation}
This represents the 6 clones of a point labeled by $(0,0)$ in $L(7,2)$. 

An important question for cosmology is ``For a given value of $p$ which values of $q$ are physically distinct?''
Roughly, two topological spaces $S_1$ and $S_2$ have ``similar shapes'' if they are homeomorphic. 
More precisely, a function $h: S_1\to S_2$ is a homeomorphism if it is continuous, one-to-one and onto, and its inverse $h^{-1}$ is also continuous.
If such a function exists then the spaces $S_1$ and $S_2$ are homeomorphic.
Two lens spaces $L(p,q)$ and $L(p',q')$ are known to be homeomorphic if and only if $p = p'$ and either $q = \pm q'\Mod{p}$ or $qq^\prime=\pm 1\Mod{p}$.

Topologically, $L(p,q)$ and $L(p,q')$ are equivalent if they are homeomorphic, resulting in catalogs of lens spaces only including one of each class.
However, a homeomorphism does not preserve all physical properties.
Thus, although spaces may ``have the same shape'', they may still be physically distinguishable by observers in those spaces.
As one example, $L(7,2)$ and $L(7,5)$ are homeomorphic since $2 = -5\Mod{7}$.
Notice that $\mat{R}_{75}$ takes $(0, 0)$ to $(1, 5)$ which is equivalent to $(1, -2)$.
Written in this more suggestive manner, if one starts at $(0,0)$ repeated application of $\mat{R}_{75}$ takes
\begin{equation}
(0,0)\to(1,-2)\to(2,-4)\to(3,-6)\to(4,-1)\to(5,-3)\to(6,-5)\to(0,0).
\label{eqn:R75-pattern}
\end{equation}
Comparing to \eqref{eqn:R72-pattern} we see that these two spaces have opposite ``handedness,'' in the sense that while $j$ steps in the same direction for both spaces, $j'$ steps in opposite directions.
Thus $L(7,2)$ and $L(7,5)$ have distinguishable clone patterns despite being homeomorphic.
More generally, this is true for all $L(p,q)$ and $L(p, p-q)$: they are homeomorphic topological spaces but have distinguishable clone patterns.
Note that for these pairs of topologies the distances between all pairs of clones will remain the same.

Continuing with $p=7$, $L(7,2)$ and $L(7,3)$ are also homeomorphic since $2\times3=6=-1\Mod{7}$.
Once again these spaces can be shown to have different clone patterns though the homeomorphism between the two spaces is less obvious.
Despite this, the distance between all pairs of clones will also remain the same.
Similarly, it can be shown that the pattern of clones seen by an observer in $L(7,2)$ is identical to that seen by some observer in $L(7,4)$ and the pattern of clones seen by an observer in $L(7,3)$ is identical to that seen by some observer in $L(7,5)$.\footnote{
      For example, the pattern of clones for $L(7,2)$ is given by \eqref{eqn:R72-pattern}, while the pattern for $L(7,4)$ is
      \[
        (0, 0) \to (1, 4) \to (2, 1) \to (3, 5) \to (4, 2) \to (5, 6) \to (6, 3) \to (0, 0).
      \]
      We see that these contain all the same clones (tuples of $j$ and $j'$) if we swap $j$ and $j'$, i.e., if we swap $(x_0, x_1)$ and ($x_2, x_3)$. (The order in the sequence is irrelevant.)  This swapping can be accomplished by the orthogonal transformation
      \[
        \mat{O} = \begin{pmatrix} \mat{0}_{2\times 2} & \identity_{2\times 2} \\
            \identity_{2\times 2} & \mat{0}_{2\times 2} \end{pmatrix}.
      \]
      Therefore, $L(7,2)$ and $L(7,4)$ have the same clone patterns.} 
Wrapping up the example, this means that there are three physically distinct lens spaces for $p=7$: $L(7,1)$ and two additional ones that can be chosen to be $L(7,2)$ and $L(7,3)$. $L(7,4)$, $L(7,5)$ and $L(7,6)$ are each physically equivalent to one of these three. 

The behavior of the $p=7$ example is generic. $L(p,q)$ is homeomorphic to $L(p,p-q)$, but they do not share the same clone pattern. There is also at most one other $0 < q'< \lfloor p/2 \rfloor$ such that $L(p,q')$ is homeomorphic to $L(p,q)$ and with its clone pattern identical to that of $L(p,p-q)$. The list of $L(p,q)$ that are distinct for cosmological purposes is thus longer than the list of $L(p,q)$ that are topologically distinct: for every topologically distinct $L(p,q)$ with $0 < q < \lfloor p/2 \rfloor$, physically one must also consider $L(p,p-q)$. If there is another homeomorphic $L(p,q')$ with $q' < \lfloor p/2 \rfloor$, then it is physically equivalent to $L(p,p-q)$.\footnote{
    Topologists are also interested in spaces that are homotopically equivalent, a weaker condition than homeomorphic: all lens spaces that are homeomorphic are also homotopically equivalent, but the converse is not true.
    The two lens spaces $L(p,q)$ and $L(p', q')$ are homotopically equivalent if and only if $p= p^\prime$ and $qq'=\pm n^2\Mod{p}$ for some $n\in\integers$.
    For example, $L(11,2)$ and $L(11,3)$ are homotopically equivalent since $2\times 3 = 6 = -4^2\Mod{11}$, but they are not homeomorphic.
    Note that $p=11$ is the smallest $p$ with both $q$ and $q'$ larger than $1$ for which there is a homotopically equivalent but not homeomorphic pair of lens spaces. 
    Homotopically equivalent lens spaces that are not homeomorphic are physically distinguishable with different clone pair separations.
}

While this holds in general, the focus of this work is on limits from circle searches which only depend on the interclone distances and not on the pattern of the clones. 
Therefore, in this work we can restrict ourselves to $0 < q < \lfloor p/2 \rfloor$ since the homeomorphic partners $L(p,p-q)$ of each lens space have the same interclone distances.

\section{Circle search for lens spaces} 
\label{sec:circle_searches}
Constraints on the non-trivial topology of the Universe can be addressed by the existing circle-in-the-sky signature searches based on CMB temperature data.
As noted above, for the lens space $L(p,q)$, the covering space contains $p$ copies of each observer, i.e., the covering space can be viewed as being tiled with each tile having a clone of each observer.
Studies originally based on the  WMAP \cite{Aurich:2013fwa} and later on higher-resolution maps from the {\it Planck} satellite \cite{Planck:2013okc} confirmed that there are no matched-circle pairs in the CMB sky maps.
This non-detection of circles in the CMB sky can be used to constrain the lens-space parameters $p$ and $q$.
A lens space can conservatively be ruled out if \emph{all observers} would see matched circle pairs.
For the inhomogeneous lens spaces (those with $q>1$) this requires comparing $\dLSS$ to the distance of every observer's nearest clone.

The distance (on the unit $S^3$) between an observer at $\vec{x}^{(0)}$ and the position of one of their clones $\vec{x}^{(j)}$ given by \eqref{eqn:cloneposn} is\footnote{
    The distance between any pair of clones $\vec{x}^{(i)}$ and $\vec{x}^{(j)}$ is
    \begin{equation*}
        d^{(p,q)}_{ij}
        = \cos^{-1}(\vec{x}^{(i)} \cdot \vec{x}^{(j)})
        = \cos^{-1}\! \left[s \cos(2\pi (i-j)/p) + (1 - s) \cos(2\pi (i-j) q/p)\right ]\,.
    \end{equation*}
    }
\begin{equation} 
     d^{(p,q)}_{j}(s) \equiv d^{(p,q)}_{0j}(s) = \cos^{-1}\! \left[ s \cos\!\left(\frac{2\pi j}{p}\right) + (1 - s) \cos\!\left(\frac{2\pi j q}{p}\right) \right]\,,
    \label{eqn:cloneseparation}
\end{equation}
for $s \equiv (x^{(0)}_0)^2 + (x^{(0)}_1)^2$.
Notice that for the homogeneous lens space $L(p, 1)$ the distance to all clones is the same for all observers
\begin{equation}
   d^{(p,1)}_j(s) = \cos^{-1}\! \left[ s \cos\!\left(\frac{2\pi j}{p}\right) + (1 - s) \cos\!\left(\frac{2\pi j }{p}\right) \right]
   = \frac{2\pi j}{p}.
    \label{eqn:dp1}
\end{equation}
On the other hand, since the lens spaces $L(p, q\neq 1)$ are globally inhomogeneous, the pattern of clones depends on the observer location.\footnote{
    The shape of the Dirichlet domain---the set of all points closer to a given observer than to any clone of that observer---also depends on the observer location for $q\neq1$.}
Thus to apply the circle search limits to a topology we must search over all observers and find the maximum distance to their nearest clone.
In other words, we must determine the maximum-minimum distance
\begin{equation}
    d^{(p,q)}_{\mathrm{max}} = \max_{0\le s \le 1}\! \left( \min_{0<j<p} d^{(p,q)}_{j}(s) \right) .
    \label{eqn:dpqmaxmin}
\end{equation}
We immediately see that in the homogeneous lens spaces 
\begin{equation}
    d^{(p,1)}_{\mathrm{max}} = \frac{2\pi}{p}.
    \label{eqn:dp1maxmin}
\end{equation}
For the inhomogeneous lens spaces ($q>1$) we compute $d^{(p,q)}_{\mathrm{max}}$ numerically.\footnote{
    Beginning from an equally spaced set of $s\in[0, 1]$, we find for each $s$ the clone $j$ to which the distance is a minimum, determine the $s$ from this set that has a maximum of these minimum distances, then repeat the process for a finer grid bracketing this $s$ until the bracket is smaller than a specified width.}

\begin{figure}
    \centering
    \includegraphics{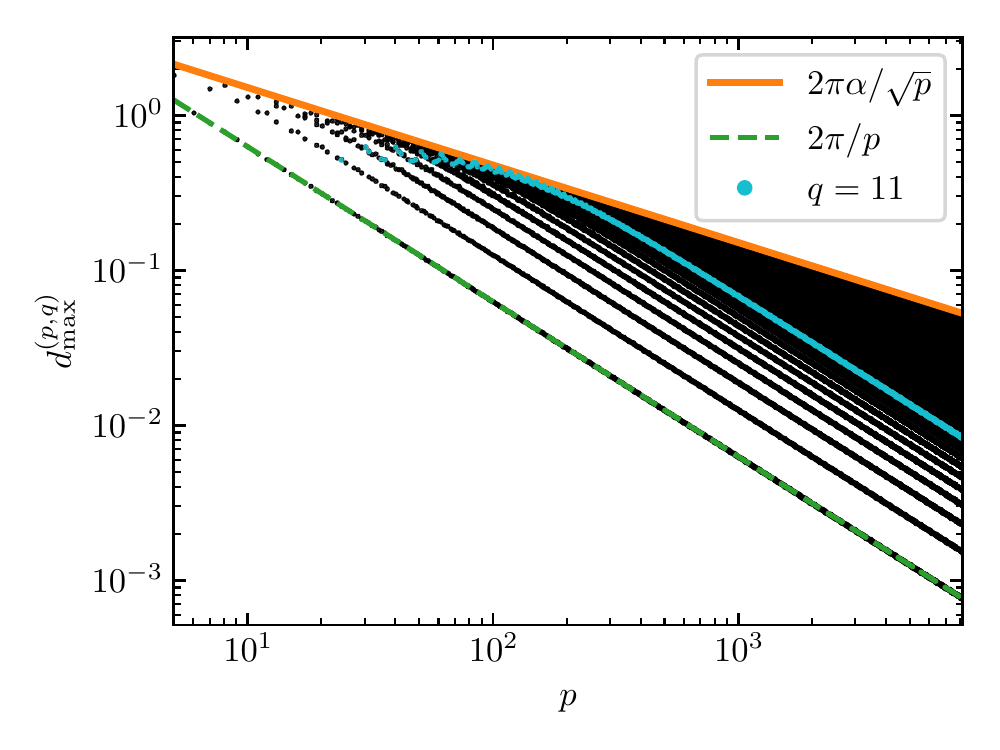}  
    \caption{The maximum-minimum distance \eqref{eqn:dpqmaxmin} for each lens space $L(p,q)$.
    Each black dot represents a different lens space.
    The smallest distance for each $p$ (lower, green, dashed line) is for the homogeneous spaces $L(p,1)$ given in \eqref{eqn:dp1maxmin}.
    The maximum for each $p$ (upper, orange, solid line) is an empirical limit \eqref{eqn:dpqmaxbound} valid for all $p<8192$.
    For illustration the distances $d^{(p,11)}_{\mathrm{max}}$ are shown in blue.
    }
    \label{fig:dpqmaxmin}
\end{figure}

The maximum-minimum distance for all lens spaces $L(p,q)$ with $p < 8192$ is shown in \cref{fig:dpqmaxmin}.
The smallest distance (lower, green, dashed line) occurs for the homogeneous lens spaces $L(p,1)$ from \eqref{eqn:dp1maxmin}.
For $q>1$ the distances appear to approach a maximum value with a shallow fall-off with respect to $p$. 
Empirically, it is found to be well approximated as
\begin{equation}
    d^{(p,q)}_{\mathrm{max}} < \frac{2 \pi \alpha}{\sqrt{p}},
    \label{eqn:dpqmaxbound}
\end{equation}
with $\alpha \simeq 0.761$. This choice for $\alpha$ is valid for all $p < 8192$, but appears to be increasing very little if at all by the time $p$ reaches this value.
This bound is shown as the upper, orange, solid line in the figure.

The scaling of this upper limit on $d^{(p,q)}_{\mathrm{max}} $ as $p^{-1/2}$ is a direct consequence of our earlier observation that  $\mat{R}^j_{pq}$ keeps both $x_0^2+x_1^2$ and $x_2^2+x_3^2$ unchanged. 
This implies that the $p$ clones of any point populate a 2-torus in $\reals^4$ (that of course lies on the $S^3$).  
If they are randomly distributed, then their average separation is proportional to $p^{-1/2}$.
An analytic  calculation of  $\alpha$ seems possible.
A first estimate is obtained by noting that the mean separation of $p$ points uniformly distributed on the maximum-area 2-torus submanifold of the 3-sphere is $d=\sqrt{8\pi/p}$, assuming that $p$ is large enough for the areas of discs of this diameter to be well-approximated by $\pi d^2/4$.  This leads to an estimate of $\alpha\simeq\sqrt{2/\pi}\simeq0.798$, which is within a few percent of the value derived numerically.

The figure also contains other noticeable band-like structure.
At first glance, since the lower bound is given by $d^{(p,1)}_{\mathrm{max}}$ it may be thought that the bands represent other fixed values of $q$.
This is not the case.
An illustrative example for $q=11$ is provided in the figure as blue dots.
Though the structure is intriguing, and more prominent as one zooms into the figure, it has no bearing on the limits presented in this work and will not be explored further here.

\section{Cosmological settings}
\label{sec:cosmology}
In order to use \eqref{eqn:dpqmaxbound} to constrain the topology of $S^3$ manifolds from cosmological data, we recall certain facts about a positively curved FLRW universe.
The spherical FLRW geometry is characterized by the locally homogeneous and isotropic FLRW metric  
\begin{equation}
    \dderiv s^2 = -c^2 \dderiv t^2 + a(t)^2 (\dderiv \chi^2 + \sin^2\chi \dderiv \Omega^2) .
\end{equation}
Here, $t$ is the cosmic time, $a(t)$ is the scale factor, $\chi$ is the comoving radial distance in units of the curvature radius $R_c$ for the 3-sphere, and $\dderiv \Omega^2 = \dderiv \theta^2 + \sin^2 \theta \, \dderiv \varphi^2$ is the infinitesimal solid angle.

The first Friedmann equation in this geometry for a universe filled with homogeneous dust of density $\rho$ (and zero pressure) and with cosmological constant $\Lambda$ is
\begin{equation}
    \label{eqn:FRW1}
    H(t)^2= \frac{8\pi G \rho(t)}{3} - \frac{c^2}{a(t)^2 R^2_c} + \frac{\Lambda c^2}{3} ,
\end{equation}
where $H(t) = \dot{a}(t)/a(t) $ is the Hubble expansion rate, $G$ is Newton's constant, and we have displayed $R_c$ explicitly.
It is convenient to rewrite this in terms of density parameters today: $\Omega_\mathrm{m}$ for matter, $\Omega_\Lambda$ for the cosmological constant, and $\Omega_K$ as an effective density parameter for curvature, all defined by
\begin{equation}
    \Omega_\mathrm{m} \equiv \frac{8\pi G \rho_0}{3 H_0^2}, \quad \Omega_\Lambda \equiv \frac{\Lambda c^2}{3H_0^2}, \quad \Omega_K \equiv -\frac{c^2}{R_c^2 H_0^2}\,.
\end{equation}
Here all quantities are written in terms of their values today at time $t_0$: $H_0 \equiv H(t_0)$, $\rho_0 \equiv \rho(t_0)$, and we have chosen $a_0 \equiv a(t_0) = 1$.
Noting that $\rho(t) = \rho_0 / a(t)^3$ for nonrelativistic matter we can, as usual, rewrite \cref{eqn:FRW1} as 
\begin{equation}
    \label{eqn:FRW1-densities}
    H^2 = \left( \frac{\dot{a}}{a} \right)^2 = H_0^2 \left( \Omega_\mathrm{m} a^{-3} + \Omega_K a^{-2} + \Omega_{\Lambda} \right) .
\end{equation}
Notice that from the definition of $\Omega_K$ the current physical curvature radius is
\begin{equation}
    R_c^{\mathrm{phys}} \equiv a_0 R_c =  \frac{c}{H_0 \sqrt{\vert\Omega_{K}\vert}} .
\end{equation}

The comoving distance between an observer and a point at redshift $z$ can be found by using $a = 1/(1+z)$ and integrating along a radial null geodesic
\begin{equation}
    \dderiv\chi = \frac{c \dderiv t}{a} = \frac{c \dderiv a}{a \dot{a}} = -c \left( \frac{a}{\dot{a}} \right) \dderiv z.
\end{equation}
Finally, using \cref{eqn:FRW1-densities}, the comoving distance 
in units of $R_c$ is given by
\begin{equation}
    \label{eqn:chi_observer}
    \chi(z) = \int \dderiv\chi = \sqrt{\vert\Omega_{K}\vert}\int^z_0 \frac{\dderiv x}{\sqrt{\Omega_\mathrm{m}(1+x)^3+ \Omega_K(1+x)^2+\Omega_{\Lambda}}} \,.
\end{equation}
As we use CMB data to constrain topology we are interested in the \emph{diameter} of the LSS, which we take to be at $z_\mathrm{LS}=1090$.
The comoving diameter of the LSS in units of $R_c$ is thus
\begin{equation}
    \label{eqn:dLSS}
    \dLSS = 2 \chi(z_{\mathrm{LS}}) . 
\end{equation}

\begin{figure}
\centering
    \includegraphics[width=3in]{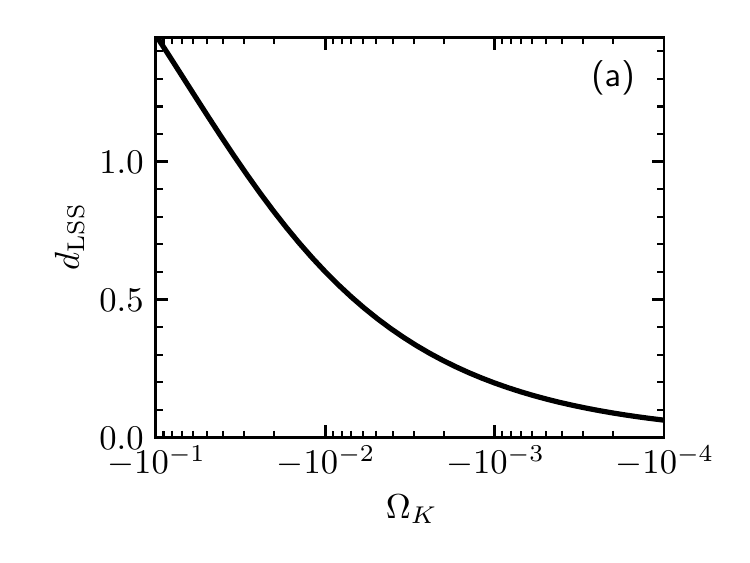} \hfill \includegraphics[width=3in]{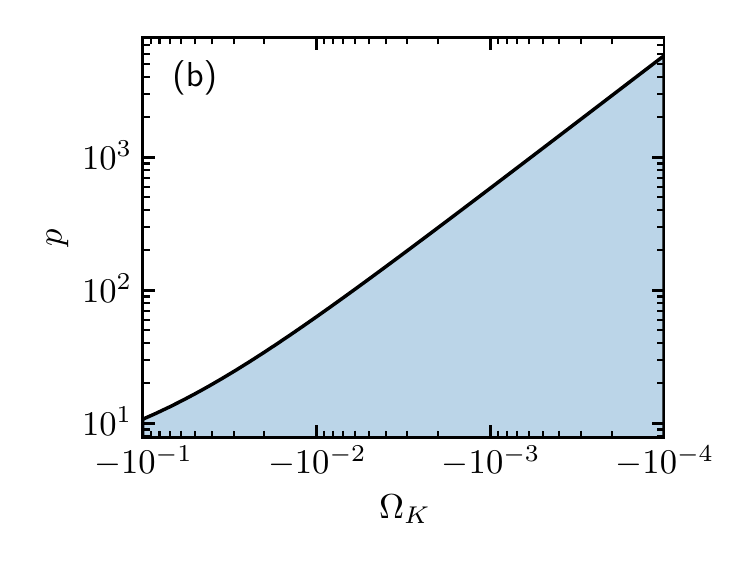}
    \caption{Implications of cosmology.
    Panel (a) on the left shows $\dLSS$ (in units of $R_c$) as a function of $\Omega_K$ for $z=1090$ and $\Omega_\mathrm{m}$ satisfying the best-fit constraint in \cref{eqn:OmegamvsOmegaK}.
    Panel (b) on the right shows how this, coupled with the non-observation of matched circles in the sky, imposes constraints on allowed lens spaces.
    The solid line is our limit $p^*$ in \eqref{eqn:pbound-ours}.
    The shaded region (below the line) shows the values of $p$ consistent with the non-observation of circles in the sky (with $f_\mathrm{O}=1$).}
    \label{fig:dLSSvsOmegaK}
\end{figure}

The value of $\dLSS$ given in \eqref{eqn:chi_observer} depends on cosmological parameters $\Omega_\mathrm{m}$ and $\Omega_K$.\footnote{Notice that the Friedmann equation \eqref{eqn:FRW1} evaluated today is $\Omega_\mathrm{m} + \Omega_\Lambda + \Omega_K = 1$, so only two of these densities are independent.}
The best-fit cosmological model from the CMB has a degeneracy between $\Omega_\mathrm{m}$ and $\Omega_K$.
To approximate this degeneracy we note from \textit{Planck} 2018 \cite{Planck:2018vyg}, Fig.~29, that the constraints on $\Omega_\mathrm{m}$ as a function of $\Omega_K$ from primary CMB anisotropies (i.e., without lensing and baryon acoustic oscillations) are quite tight, allowing us to find an empirical relationship from the central contour in the $\Omega_\mathrm{m}$-$\Omega_K$ plane,
\begin{equation}
    \label{eqn:OmegamvsOmegaK}
    \Omega_\mathrm{m} \simeq 0.314 - 3.71\Omega_K.
\end{equation}
Based on this relation,  $\dLSS$ versus $\Omega_K$  in units of $R_c$ is calculated and shown in \cref{fig:dLSSvsOmegaK}. 
However, it should also be noted that both the location of the central contour and the tightness of the fit to \eqref{eqn:OmegamvsOmegaK} were obtained in the context of  the specific power spectra adopted by {\it Planck}, which may very well require modification in the context of non-trivial topology, especially on large scales.

\section{Cosmological constraints on detectability of lens spaces}
\label{sec:constraints}
To combine the previous two sections, we use the condition that all observers in a lens space will see circles in the sky when
\begin{equation}
    \label{eqn:dpqltfOdLSS}
    d^{(p,q)}_{\mathrm{max}} < f_\mathrm{O} \dLSS.
\end{equation}
$f_\mathrm{O} d_\mathrm{LSS}$ is the observational lower limit, from the CMB, on the length of the shortest closed spatial geodesic.
As discussed in \cref{sec:introduction}, $f_\mathrm{O}\simeq 0.985$ at 95\% confidence level, but is not much smaller at much higher confidence \cite{Vaudrevange:2012da}. 
Coupled with the upper limit on $d^{(p,q)}_{\mathrm{max}}$ from \eqref{eqn:dpqmaxbound}, this can be used to rule out lens spaces with too small distances to the nearest clone.
Every observer in the lens space $L(p,q)$ will see circles if
\begin{equation}
    \frac{2\pi\alpha}{\sqrt{p}} < f_\mathrm{O} \dLSS.
\end{equation}
Solving for $p$ and inverting the logic, the only lens spaces consistent with the non-detection of matched circles in the sky have
\begin{equation}
    p < \frac{4\pi^2 \alpha^2}{f_\mathrm{O}^2\dLSS^2} \equiv p^*.
    \label{eqn:pbound-ours}
\end{equation}
We shall take $f_\mathrm{O}=1$ for illustrative purposes in our figures, since our topology limits are already subject to some uncertainty due to their dependence on cosmological parameters.
The value of $\dLSS$ depends on cosmological parameters (e.g., \cref{eqn:chi_observer,eqn:dLSS}) and can be reduced to a function of $\Omega_K$ using the approximation in \cref{eqn:OmegamvsOmegaK}.
The resulting dependence is shown in the left panel (a) of \cref{fig:dLSSvsOmegaK}.
It thus follows that the bound set by $p^*$ is also a function of $\Omega_K$.
The parameter space of lens spaces as a function of $\Omega_K$ is shown in the right panel (b) of \cref{fig:dLSSvsOmegaK} with the bound from \eqref{eqn:pbound-ours} as a consequence of the absence of matched circles in the CMB.
The white region represents the region of the parameter space $(\Omega_K, p)$ excluded for any valid choice of $q$, while the shaded region represents the allowed parameters for some $q$.
Recall that precise limits on $(p,q)$ depend on the specific values of cosmological parameters.

We note that the $(p,q)$ parameter space allowed by the lack of matched circles is still large given that our current knowledge of $\Omega_K$ is poor. 
In particular, the closer $\Omega_K$ is to zero, the weaker are the constraints on $p$ and $q$. Of course, there is more cosmological information available in the CMB than just whether or not there are matched circles; we will study the broader effects of topology on CMB anisotropies in $S^3$ with non-trivial topology in future papers.

As noted in the introduction, observational constraints on lens spaces had been considered in \rcite{Gomero:2001gq}, where the authors demonstrated that, for a given $(p,q)$, the absolute (``global'') maximum distance to the nearest clone is $2\pi q/p$ in units of $R_c$.  
(This is realized at $s=0$ for the $j=1$ clone, as seen from \cref{eqn:cloneseparation}.)
When applied to the LSS, and given the non-observation of matched circles, this provides an alternate  $q$-dependent upper bound on allowed lens spaces, requiring
\begin{equation}
    \frac{2\pi q}{p} < f_\mathrm{O} \dLSS\,.
\end{equation}
This translates into the upper bound
\begin{equation}
    p < \frac{2\pi q}{f_\mathrm{O} \dLSS} \equiv p^*_q.
    \label{eqn:pbound-q}
\end{equation}
The tightest constraint on lens spaces is determined from a combination of the two bounds \eqref{eqn:pbound-ours} and \eqref{eqn:pbound-q}:
no observer will see circles when
\begin{equation}
    \min\!\left( \frac{2\pi\alpha}{\sqrt{p}}, \frac{2\pi q}{p} \right) > f_\mathrm{O}\dLSS,
    \label{eqn:dLSS_limit}
\end{equation}
where the first distance corresponds to the bound $p^*$ and the second to $p^*_q$.

From this we see that $p^*$ in \eqref{eqn:pbound-ours} provides a stricter bound than $p^*_q$ in \eqref{eqn:pbound-q} when
\begin{equation}
    q > \alpha \sqrt{p}.
    \label{eqn:pbound-transition}
\end{equation}

\begin{figure}
    \centering
    \includegraphics[width=\textwidth]{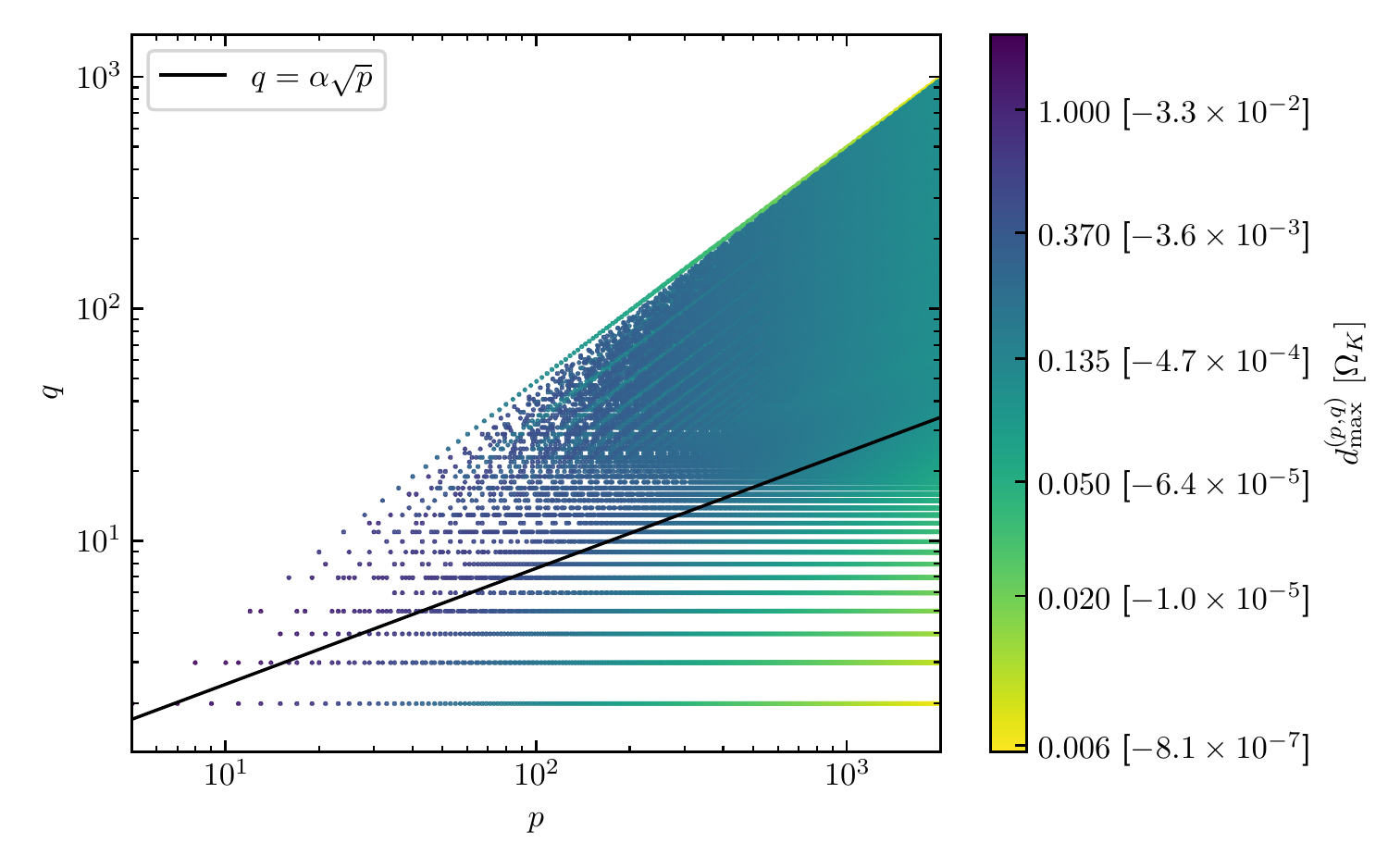}
    \caption{The maximum-minimum clone separation  in units of $R_c$, $d_{\mathrm{max}}^{(p,q)}$ (from \cref{eqn:dpqmaxmin}), for all lens spaces $L(p,q)$ with $p<8192$.
    The color represents the magnitude of the distance in units of the curvature radius $R_c$.
    This distance has been converted (for the empirical relationship \eqref{eqn:OmegamvsOmegaK}) to a  minimum allowed value of $\Omega_{K}$ (in square brackets)  by equating $d_{\mathrm{max}}^{(p,q)}$ to $\dLSS$, which is the minimum interclone distance given the non-observation of matched circles in the CMB sky.
    }
    \label{fig:distance_pq}
\end{figure}

The connections between the maximum-minimum distance and cosmological data, as embodied in the curvature through $\Omega_K$, for each lens space $L(p,q)$ is summarized in \cref{fig:distance_pq}.
The color represents the minimum value of $\dLSS$ for which some observers in $L(p,q)$ would see circles.
For any value of $\dLSS$ less than this, some observers would see circles.
For any value of $\dLSS$ larger than this, \emph{no observers would see circles}.
The black line separates the space into regions where the bound from \rcite{Gomero:2001gq} in \eqref{eqn:pbound-q} on the distance is tighter (below the line) and the bound from this work in \eqref{eqn:pbound-ours} is tighter (above the line), with the transition happening when $q=\alpha \sqrt{p}$.
This result is independent of cosmology.

Additionally, the color can also be interpreted as giving a lower limit on the value of $\Omega_K$ (given in square brackets) for each lens space consistent with the non-detection of pairs of matched circles in the CMB sky.
The quoted cosmology-dependent value of $\Omega_K$ is computed by requiring $d_{\mathrm{max}}^{(p,q)} = \dLSS(\Omega_K)$ 
for the fit \eqref{eqn:OmegamvsOmegaK} to \textit{Planck} 2018 cosmological data, as shown in panel (a) of \cref{fig:dLSSvsOmegaK}.
Again we stress that the $\dLSS$ limit \eqref{eqn:dLSS_limit} is independent of the specific values of cosmological parameters (and the details of power spectrum), whereas the $\Omega_K$ limit displayed in Fig.~\ref{fig:distance_pq} is not.

\section{Conclusions}
\label{sec:conclusions}
The local geometry of the Universe is nearly flat, which means that it is consistent with a small, positive (or negative), isotropic 3-curvature  \cite{Anselmi:2022uvj}.
However, positive curvature does not imply that the topology is necessarily that of the covering space of spherical geometry, $S^3$.
The 3-sphere admits a countable infinity of other topologies, among them the lens spaces $L(p,q)$, which are quotients of $S^3$ by $\integers_p$, the cyclic group of order $p$. The integer parameter $q$ (with $0<q<p$) indexes different realizations of such quotients (cf., Eq.~\eqref{eqn:Rjpq}), though not all values of $q$ in this range give manifolds, and not all values that do give manifolds are  distinct, either topologically or physically.  

All manifolds with $S^3$ local geometry, and in particular all lens spaces, are compact, and the larger $p$ is, the smaller the volume of the space for fixed curvature radius $R_c$. Meanwhile, independent analyses of both WMAP and {\it Planck} temperature data have shown us that the shortest closed distance around the Universe through our location is greater than $f_\mathrm{O}=98.5\%$ of the diameter of the last-scattering surface, $d_\mathrm{LSS}$
(as remarked above, we take $f_\mathrm{O}=1$ rather than $0.985$ for simplicity).
For fixed values of  $R_c$, this places an upper limit on $p$ and more specifically restricts the values of $(p,q)$  according to \eqref{eqn:dLSS_limit}. 
We have shown that this limit on $(p,q)$ is considerably more stringent than the previous limit \eqref{eqn:pbound-q} for most values of $(p,q)$. In Fig.~\ref{fig:distance_pq}, we have presented this limit, giving cosmology-independent values of the maximum-minimum clone separation  in units of $R_c$, $d_{\mathrm{max}}^{(p,q)}$, and (in square brackets) cosmology-dependent values of the maximum allowed value of $\Omega_K$ given the empirical relation \eqref{eqn:OmegamvsOmegaK} obtained from the primary CMB anisotropies measured by \textit{Planck} \cite{Planck:2018vyg}.

Future work will extend these limits to the other $S^3$ topologies: the prism manifolds (a.k.a.\ dihedral spaces), and the tetrahedral, octahedral, and icosahedral spaces.
We will also present the spin-0 and spin-2 eigenmodes of spherical geometries, and the correlation matrices of density and CMB fluctuations of all types, allowing us to predict the statistical properties of the CMB and of other cosmological observables in all $S^3$ manifolds. 

\acknowledgments
S.S., C.J.C., and G.D.S. thank D. Singer for extended conversations about the geometry and topology of $S^3$ manifolds.  
We thank J. Weeks for several valuable conversations.
This work made use of the High-Performance Computing Resource in the Core Facility for Advanced Research Computing at Case Western Reserve University. Y.A. acknowledges support by the Spanish Research Agency (Agencia Estatal de Investigaci\'on)'s grant RYC2020-030193-I/AEI/10.13039/501100011033, by the European Social Fund (Fondo Social Europeo) through the  Ram\'{o}n y Cajal program within the State Plan for Scientific and Technical Research and Innovation (Plan Estatal de Investigaci\'on Cient\'ifica y T\'ecnica y de Innovaci\'on) 2017-2020, and by the Spanish Research Agency through the grant IFT Centro de Excelencia Severo Ochoa No CEX2020-001007-S funded by MCIN/AEI/10.13039/501100011033.  T.S.P. acknowledges financial support from the Brazilian National
Council for Scientific and Technological Development (CNPq) under grants 312869/2021-5
and 88881.709790/2022-0.
C.J.C., G.D.S., A.K., and D.P.M.\ acknowledge partial support from NASA ATP grant RES240737; G.D.S. and Y.A.\ from the Simons Foundation; G.D.S. and A.S.\ from DOE grant DESC0009946; G.D.S., Y.A., and A.H.J.\ from the Royal Society (UK); and A.H.J.\ from STFC in the UK\@.
A.T.\ is supported by the Richard S.\ Morrison Fellowship.
G.D.S. and Y.A. thank the INFN (Sezione di Padova), and G.D.S., S.A., D.P.M., and A.T. thank the IFT for hospitality where part of this work was accomplished.

\appendix

\label{app:generators}

\bibliographystyle{utphys}
\bibliography{topology,additional}
\end{document}